\documentclass[%
 reprint,
 amsmath,amssymb,
 aps,
prl,
]{revtex4-1}

\usepackage{graphicx}
\usepackage{dcolumn}
\usepackage{bm}
\begin{document}

\preprint{Mn3Sn S}

\title{Anomalous Nernst and Righi-Leduc effects in Mn$_{3}$Sn: Berry curvature and entropy flow}

\author{Xiaokang Li$^{1}$, Liangcai Xu$^{1}$, Linchao Ding$^{1}$, Jinhua Wang$^{1}$, Mingsong Shen$^{1}$, Xiufang Lu$^{1}$, Zengwei Zhu$^{1, *}$ and Kamran Behnia$^{1,2,*}$}

\affiliation{(1) Wuhan National High Magnetic Field Center\\
 School of Physics, Huazhong University of Science and Technology,  Wuhan  430074, China\\
(2)Laboratoire de Physique Et d'Etude des Mat\'{e}riaux (UPMC-CNRS),ESPCI Paris, PSL Research University\\
75005 Paris, France\\
}

\date{June 16, 2017}

\begin{abstract}
We present a study of electric, thermal and thermoelectric response in noncollinear antiferromagnet Mn$_{3}$Sn, which hosts a large Anomalous Hall Effect (AHE). Berry curvature generates off-diagonal thermal(Righi-Leduc) and thermoelectric(Nernst) signals, which are detectable at room temperature and invertible with a small magnetic field. The thermal and electrical Hall conductivities respect the Wiedemann-Franz law, implying that the transverse currents induced by Berry curvature are carried by Fermi surface quasi-particles. In contrast to conventional ferromagnets, the anomalous Lorenz number remains close to the Sommerfeld number over the whole temperature range of study, excluding any contribution by inelastic scattering and pointing to Berry curvature as the unique source of AHE. The anomalous off-diagonal thermo-electric and Hall conductivities are strongly temperature-dependent and their ratio is close to k$_{B}$/e.
\end{abstract}

\maketitle

The ordinary Hall effect, the transverse electric field generated by a longitudinal charge current in presence of a magnetic field, is caused by the Lorentz force exerted by magnetic field on charge carriers.  In ferromagnetic solids, there is an additional component to this response (known as extraordinary or anomalous), thought to arise as a result of a sizeable magnetization. During the past decade, a clear link  between the Anomalous Hall Effect (AHE) and the Berry curvature of Bloch waves has been established\cite{Nagaosa2010,Xiao2010}. Charged carriers of entropy are also affected by the Lorenz force. Therefore, one expects a transverse component to thermal conductivity called the Righi-Leduc (or the thermal Hall) effect\cite{Zhang2000} in presence of magnetic field. This is also the case of thermoelectric conductance, which acquires an off-diagonal component, $\alpha_{ij}$, intimately linked to the Nernst coefficient, directly measurable by experiment\cite{Behnia2016}. When the Berry curvature replaces the magnetic field, counterparts of the AHE appear in the thermal and thermoelectric response of ferromagnets\cite{Lee2004,Miyasato2007,Xiao2006,Onose2008}. They can be an additional source of information regarding the fundamental mechanism leading to the generation of dissipationless transverse currents. Recently,  following a proposition by Chen, Niu and Macdonald\cite{Chen2014},  Nakasutji \emph{et al.} and Nayak \emph{et al.} found a large AHE in  Mn$_{3}$Sn\cite{Nakatsuji2015} and Mn$_{3}$Ge\cite{Nayak2016,Kiyohara2016}, which are noncollinear antiferromagnets at room temperature. Several recent theoretical studies were devoted to this issue\cite{Kubler2014,Zhang2016,Suzuki2016}.

In this letter, we present a study of  Anomalous Righi-Leduc and Nernst effects (ANE) in Mn$_{3}$Sn in order to quantify the amplitude of these coefficients compared to their Hall counterpart. We detect a large  Anomalous Righi-Leduc conductivity and find that its magnitude corresponds to what is expected according to the Wiedemann-Franz (WF) Law over an extended temperature window. The result confirms a theoretical prediction by Haldane\cite{Haldane2004} with important consequences for the debate regarding the two alternative formulations of anomalous Hall effect\cite{Haldane2004,Chen2013,vanderblit2014}. AHE can be formulated as  a property of the whole Fermi sea or the Fermi surface\cite{Haldane2004,vanderblit2014}. The expected magnitude of the Anomalous Hall Conductivity (AHC) in both pictures is identical (as explicitly shown in the case of BCC iron\cite{Yao2004,Wang2006,Gosalbez2015}). The verification of the Wiedemann-Franz law allows for an experimental distinction, since the validity of this law is straightforward in the Fermi-surface picture, but not necessarily in the Fermi-sea picture. We compare the robustness of the WF law in Mn$_{3}$Sn and in Fe and Ni-based ferromagnets\cite{Onose2008} and find that inelastic scattering does not play any detectable role in generating the anomalous transverse coefficients of Mn$_{3}$Sn, in contrast to common ferromagnets.  We also quantified the anomalous transverse thermoelectric response, $\alpha^{A}_{ij}$, and argue that the magnitude and the temperature dependence of the ratio of $\sigma^{A}_{ij}$ to $\alpha^{A}_{ij}$ is a source of information on the location of that Weyl nodes.

The  Mn$_{3}$Sn single crystal was grown using the Bridgman-Stockbarger technique. Thermal and thermoelectric conductivity  was measured with  thermocouples. (See\cite{supplement}for details). The temperature dependence of the longitudinal transport coefficients are shown in Fig.1. As seen in the figure, resistivity attains a magnitude of 250 $\mu\Omega$ cm and becomes almost flat at room temperature. Electric and thermal conductivities are both almost isotropic, in contrast to the Seebeck coefficient. The temperature dependence of magnetization, measured in Field-Cooled (FC) conditions, is similar to what was previously reported\cite{Ohmori1987}. It shows that the system becomes magnetically ordered below T$_{N}$=420 K and above T$_{1}$=200 K with a weak ferromagnetic remanence. Tomiyoshi and Yamauguchi\cite{Tomiyoshi1982} studied the magnetic  texture of this system by polarized neutron diffraction decades ago and following an earlier study\cite{Zimmer1973}, suggested a triangular spin structure stabilized by the Dzyaloshinski-Moriya interaction [Fig. 1e]. An alternative magnetic texture [Fig. 1f] will be considered below in the light of a recent theoretical calculation\cite{Zhang2016}.

\begin{figure}
\includegraphics[width=9cm]{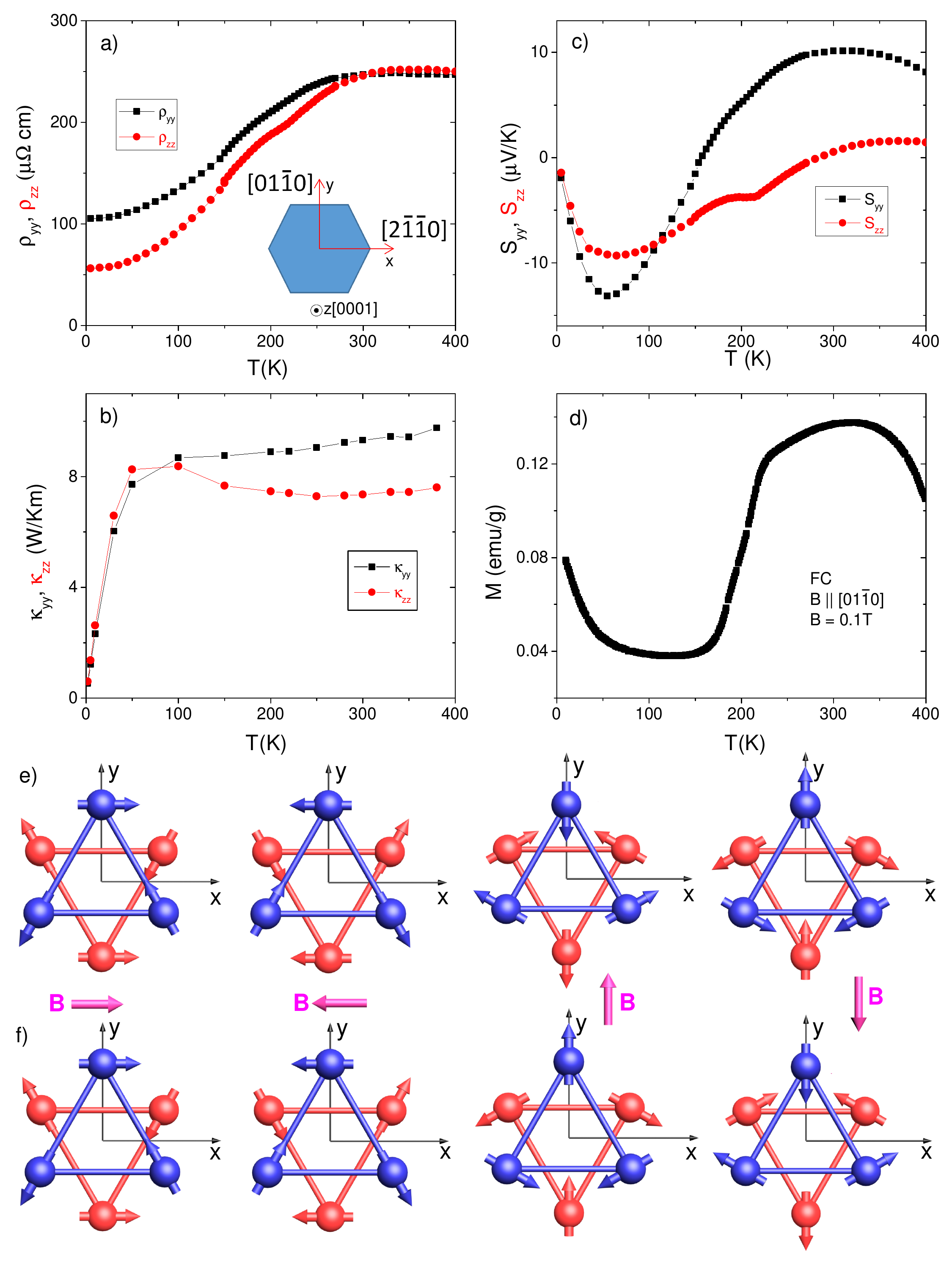}
\caption{(Color online) Zero-field temperature dependence of resistivity a), thermal conductivity b), Seebeck coefficient c) and the field cooled (FC) magnetization d). Panels e) and f) show magnetic texture for four different orientations of magnetic field. Blue and red circles represent Mn atoms in adjacent planes. Panel e) is the set of configurations proposed in ref.\cite{Tomiyoshi1982}. Panel f) starts with the chirality considered in a theoretical calculation\cite{Zhang2016} and assumes that the spins rotate freely with the magnetic field. In e),  Mn spins, which are parallel to the field when B$\|$x, become anti-parallel when B$\|$y. In f), the magnetic field and the spins keep their mutual orientations.}
\end{figure}

Two points are to be noticed about transport of charge and entropy in this metal. First, given the carrier density (n=2$\times$10$^{22}$cm$^{-3}$\cite{Nakatsuji2015,supplement}), a room-temperature resistivity of 250 $\mu\Omega cm$ implies a mean-free-path as short as 0.7 nm. This is comparable to the lattice parameters (a=0.566nm and c=0.453nm\cite{Tomiyoshi1982}). Thus, the system is close to the Mott-Ioffe-Regel limit and the saturation of resistivity between 300 K and 400 K is not surprising\cite{Gunnarsson2003}. Second, the room-temperature Lorenz number ($L_{ii}=\frac{\kappa_{ii}(300K) \rho_{ii}(300K)}{300K}$) is almost twice the Sommerfeld number, L$_{0}$= 2.44$\times$ 10$^{-8}$V$^{2}$K$^{-2}$. Therefore, the total contribution of phonons and  magnons to the longitudinal thermal transport is comparable to the electronic heat transport.

\begin{figure}
\includegraphics[width=9cm]{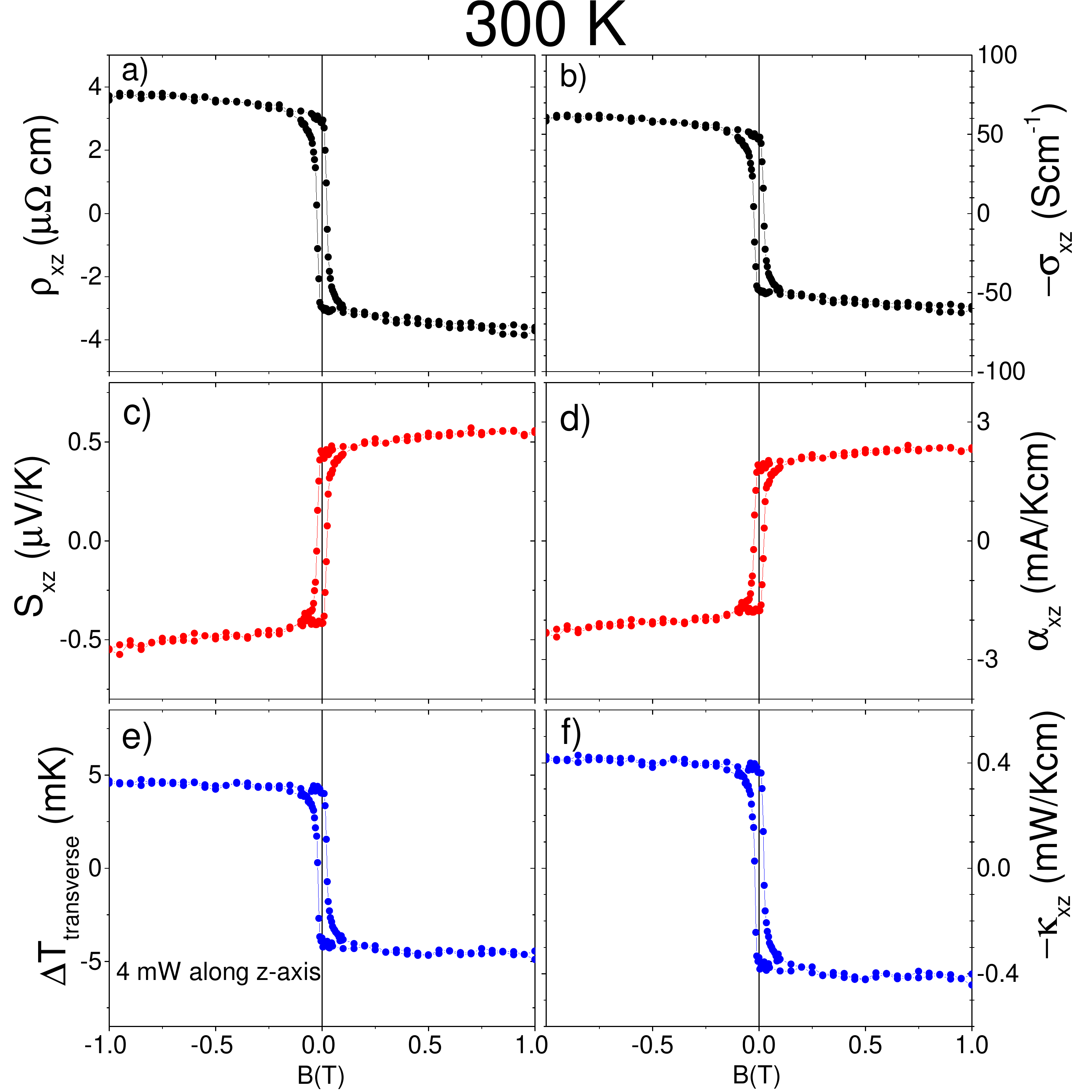}
\caption{(Color online) Room-temperature field dependence of  transport coefficients with magnetic field is applied along the y-axis ([01\={1}0]) and the charge or heat current along the z-axis ([0001]). a) Hall resistivity, $\rho_{xz}$); b)Hall conductivity, $\sigma_{xz}$, extracted from $\rho_{xz}$, $\rho_{xx}$ and $\rho_{zz}$; c) Nernst signal, S$_{xz}$; d) Transverse thermoelectric conductivity, $\alpha_{xz}$, extracted from S$_{xz}$, S$_{zz}$, $\rho_{xx}$, $\rho_{zz}$ and $\rho_{xz}$; e) The transverse thermal gradient generated by a longitudinal heat current along the z-axis. f) The extracted Righi-Leduc coefficient, $\kappa_{xz}$.}
\end{figure}

What makes this magnetic metal remarkable is its transverse transport, illustrated in Fig.2. With a current along [0001] (dubbed z-axis and) a magnetic field  along [01\={1}0] (dubbed y-axis), there is a large and hysteretic jump in the Hall resistivity, $\rho_{xz}$. Its magnitude (7.6 $\mu \Omega cm$) is comparable to what was reported previously\cite{Nakatsuji2015} and is reversible with a field as small as 0.1 T. The Nernst coefficient (Fig. 2c) shows a  jump of 1 $\mu$V/K, a remarkably large value.  Given the low mobility ($\sim$ 1.7 cm$^{2}$V$^{-1}$s$^{-1}$) and the large Fermi energy ($\sim$ 2.6eV) of the system, this is six orders of magnitude larger than the expected quasi-particle response\cite{Behnia2016}. The same set-up was used to measure the AHE and ANE of an iron  single crystal to obtain data similar to what was previously reported\cite{Dheer1967,Watzman1967}. They are presented in Fig. S1 of the supplement\cite{supplement}. In Mn$_{3}$Sn, the anomalous transport coefficients dominate their ordinary counterparts, leading to a step-like profile quite distinct from what is seen in ferromagnets such as iron\cite{Dheer1967,Watzman1967} or cobalt\cite{kotzler2005}.

\begin{figure}
\includegraphics[width=9cm]{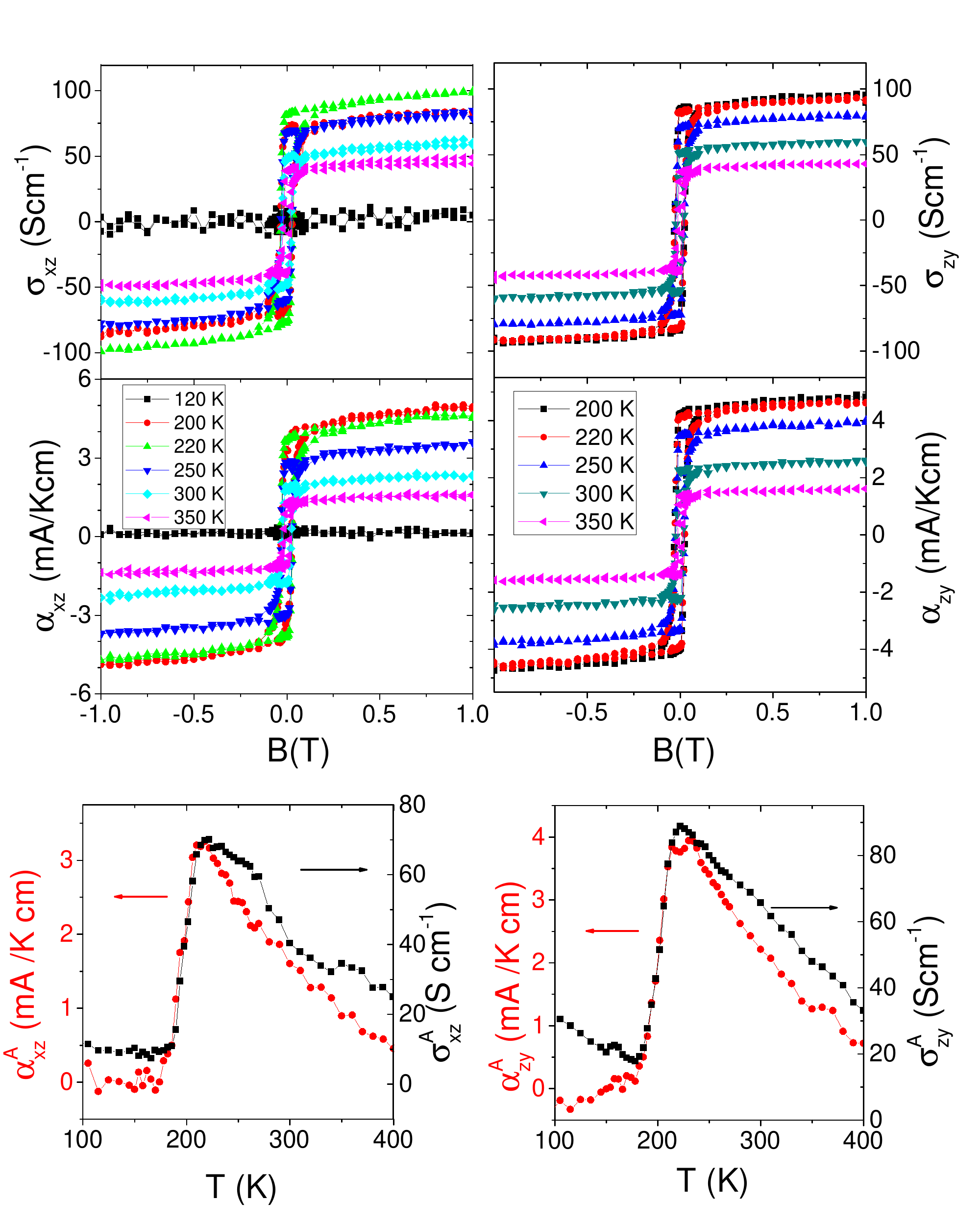}
\caption{(Color online) Top: Field dependence of $\sigma_{xz}$ and $\alpha_{xz}$ (left) and $\sigma_{zy}$  and  $\alpha_{zy}$(right) at different temperatures. The spin magnetic structure is sketched in the inset. Bottom: The temperature dependence of $\sigma^{A}_{xz}$ and $\alpha^{A}_{xz}$ (left) and $\sigma^{A}_{zy}$  and  $\alpha^{A}_{zy}$(right). Note the phase transition at T$^{*} \sim $ 200 K leading to the collapse of all anomalous transport coefficients. For both orientations,$\alpha^{A}_{ij}$ raises faster than $\sigma^{A}_{ij}$.  }
\end{figure}

From Hall resistivity, $\rho_{xz}$ and Nernst response, $S_{xz}$, one can extract the magnitude of Hall conductivity, $\sigma_{xz}$ (Fig. 2b), and transverse thermoelectric conductivity, $\alpha_{xz}$ (Fig. 2d). This can be done by manipulating $\overline{\rho}$, $\overline{\sigma}$, $\overline{\alpha}$ and $\overline{S}$ tensors(See the supplement\cite{supplement}). We also found a purely thermal counterpart to anomalous electric and thermoelectric responses. As seen in Fig.2e, applying a thermal current  along the z-axis generates a transverse thermal gradient along the x-axis, which can be inverted by a magnetic field applied along the y-axis. This allows to extract the amplitude of thermal Hall (Righi-Leduc) conductivity, $\kappa_{xz}$ (Fig. 2f).

The anomalous electric and thermoelectric transport coefficients were studied at different temperatures in two configurations (B//y and B//x). Fig. 3 presents the field dependence of $\sigma_{xz}$,  $\sigma_{zy}$, $\alpha_{xz}$ and  $\alpha_{zy}$ at several temperatures. As seen in the figure, the behavior remains step-like down to 200K and the magnitude of all coefficients smoothly increases with decreasing temperature. At T$_{1}\simeq$ 200K, the triangular spin order is destroyed\cite{Kren1975,Tomiyoshi1986,Brown1990} and all anomalous transport coefficients  disappear below this temperature as seen in the bottom panels of Fig. 3, which present the temperature dependence of  $\sigma^{A}_{ij}$ and $\alpha^{A}_{ij}$.  This confirms that the remarkably large magnitude of zero-field transverse transport coefficients is a property of the non-collinear antiferromagnet stabilized between 420 K and 200K. This phase transition to what seems to be a glassy ferromagnetic order\cite{Feng2006,Brown1990} at 200 K is specific to Mn$_{3}$Sn and does not occur in its sibling Mn$_{3}$Ge\cite{Nayak2016,Kiyohara2016}. In the following, we will focus on the magnitude of the anomalous transport coefficients in the magnetic phase hosting the triangular spin order. Two correlations are relevant here. The first is the Wiedemann-Franz law:

 \begin{equation}\label{1}
L^{A}_{ij}= \frac{\kappa^{A}_{ij}}{T \sigma^{A}_{ij}} = L_{0}=\frac{\pi^{2}}{3}(\frac{k_{B}}{e})^2
\end{equation}

The second is the Mott formula:
\begin{equation}\label{2}
\alpha_{ij}= -\frac{\pi^{2}}{3}\frac{k_{B}}{e}k_{B}T\frac{\partial\sigma_{ij}}{\partial E}|_{E_{F}}
\end{equation}

Fig. 4 presents the field dependence of $\kappa_{xz}$ and  $\kappa_{zy}$ at different temperatures and the temperature dependence of anomalous Lorenz number, L$^{A}_{ij}=\kappa^{A}_{ij} /T \sigma^{A}_{ij}$. As one can see in the bottom panels of Fig. 4, in the temperature range of triangular spin order (i.e. from 400 K down to 200 K),  the experimentally-resolved  $L^{A}_{ij}$ remains close to $L_{0}$. Thus, the anomalous  thermal and electrical Hall conductivities obey the correlation dictated by the Wiedemann-Franz law.

The intrinsic AHE arises as a result of an additional term in the group velocity of Bloch electrons\cite{Chang1996}:
\begin{equation}\label{3}
\dot{r}=\frac{1}{\hbar}\frac{\partial\epsilon_{n(\mathbf{k})}}{\partial\mathbf{k}}+\mathbf{\dot{k}}\times\Omega_{n}(\mathbf{k})
\end{equation}

Here, $\Omega_{n}(\mathbf{k})$ is the local Berry curvature. In presence of an electric field, $E$,  $\mathbf{\dot{k}}=-e E/\hbar$. Starting from this, an expression for $\sigma^{A}_{ij}$  in presence of  Berry curvature, $\Omega^{k}_{n}$ (n is a band index) has been driven\cite{Xiao2006,Nagaosa2010,Xiao2010}:

\begin{equation}\label{4}
\sigma^{A}_{ij}=\frac{-e^{2}}{\hbar}\sum_{n}\int_{BZ}\frac{d^{3}k}{(2\pi)^{3}}f_{n}(k)\Omega^{k}_{n}(k)
\end{equation}

The expression is similar to the topological formulation of the Quantum Hall effect\cite{Kohmoto1985}. Note that the integral is taken over the whole Brillouin zone (BZ). Haldane\cite{Haldane2004} noticed that in such an expression, AHE appears as a property of the whole Fermi sea and not the Fermi surface. But, this would contradict the sprit of Landau's Fermi liquid theory. He showed that an alternative expression for AHE is\cite{Haldane2004,vanderblit2014}:
\begin{equation}\label{5}
\sigma^{A}_{ij}=\frac{-e^{2}}{\hbar}\sum_{n}\int_{S_{n}}\frac{d^{2}k}{(2\pi)^{2}}[\Omega^{k}_{n}(k).\hat{n}(k)]\mathbf{k}
\end{equation}

Here, the integral is taken over the Fermi surface S$_n$ of band n and $\hat{n}(k)$ is the unit vector perpendicular to the surface. The two formulations were recognized as equiavelent\cite{Nagaosa2010} and yield quasi-identical values of $\sigma^{A}_{xy}$ in  iron ($\sim750 Scm^{-1}$)\cite{Gosalbez2015,Yao2004},  close to the room-temperature experimental value at ($\sim1000 Scm^{-1}$\cite{Dheer1967,supplement}).

Let us consider what the existence of $\kappa^{A}_{ij}$ implies. If the electronic states in question possess an entropy, $S_k$, a temperature gradient generates a force and  $\mathbf{\dot{k}}=-S_k\nabla T/\hbar $. Now, in a Fermi-Dirac distribution, the interface between occupied and unoccupied states is the only reservoir of entropy\cite{Behnia2015}. In other words, only states at the Fermi surface (and not those deep inside the Fermi sea) can feel a force exerted by a temperature gradient. The validity of the Wiedemann-Franz law implies that the transverse electronic flow generated by Berry curvature is a flow of charge and entropy with a ratio of $\frac{\pi^{2}}{3}(\frac{kB}{e})^2$. This rules out any contribution to $\sigma^{A}_{ij}$ by Weyl nodes of the entirely occupied bands, the central issue in the Fermi-sea \emph{vs.} Fermi-surface debate\cite{Chen2013,vanderblit2014}.

Previous to this study, Onose \emph{et al.}\cite{Onose2008} measured the anomalous thermal Hall conductivity of ferromagnetic Ni and Ni$_{0.97}$Cu$_{0.03}$ and found that the Wiedemann-Franz law is satisfied at low temperatures. We performed similar measurements on a Fe single crystal\cite{supplement} and found similar results. The lower panels of Fig.4 compare all sets of data. One can see that in common ferromagnets, $L^{A}_{ij}$ is close to L$_{0}$ in the zero-temperature limit and a steady downward trend is visible at finite temperature. Now, Wiedemann-Franz law is only valid in the absence of inelastic scattering\cite{Ziman}. The Lorenz-Sommerfeld  ratio is expected to deviate downward from unity in presence of inelastic scattering. Such a deviation is clearly present in the Ni and Ni$_{0.97}$Cu$_{0.03}$\cite{Onose2008} as well as Fe, but undetectable in Mn$_{3}$Sn. This implies that inelastic scattering plays a role in generating the room-temperature anomalous transverse response in the ferromagnets, but not in Mn$_{3}$Sn.  Therefore, one can  exclude not only a phonon contribution\cite{Strohm2005} but also skew scattering by magnons as a partial source of AHE in Mn$_{3}$Sn. Our study therefore identifies the robustness of the WF law as a mean to distinguish between distinct sources of AHE\cite{Nagaosa2010}.

\begin{figure}
\includegraphics[width=9cm]{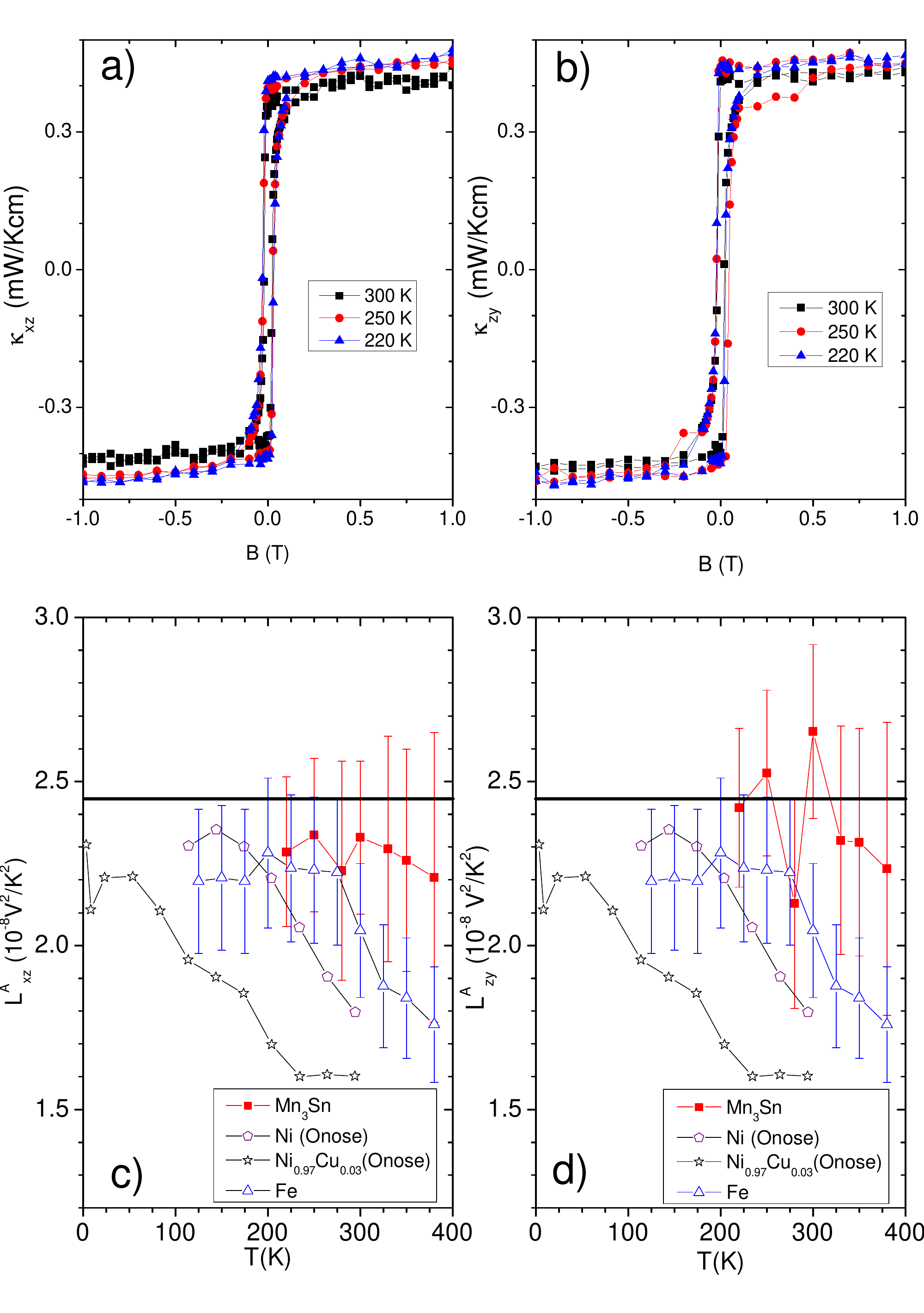}
\caption{(Color online)- Top: The field-dependence of $\kappa_{xz}$ a) and  $\kappa_{zy}$ b) at different temperatures (More data at other temperatures are in the supplement\cite{supplement}). Bottom:  The temperature dependence of anomalous Lorenz number $L^{A}_{xz}$ c) and  $L^{A}_{zy}$ d). The anomalous Lorenz number of Ni\cite{Onose2008} and Ni$_{0.97}$Cu$_{0.03}$\cite{Onose2008} and Fe\cite{supplement} are also shown. Error bars are large, because the thermal and electrical measurements were performed with different contacts. Solid horizontal lines represent L$_0$.}
\end{figure}

The peak magnitude of AHC found in this study ($\sigma^{A}_{xz}\simeq 70 Scm^{-1}$ $\sigma^{A}_{zy}\simeq 90 Scm^{-1}$ at 200K) is close to what was reported previously\cite{Nakatsuji2015}. First principle theory has shown that the expected magnitude of AHC strongly depends on the chirality of the spin texture\cite{Kubler2014}. According to a recent theoretical study\cite{Zhang2016}, for the spin texture shown in Fig. 1f, $\sigma^{A}_{xz}= 133 Scm^{-1}$ and $\sigma^{A}_{zy}=0$. However, if the spin structure rotates easily with the magnetic field, then one expects comparable magnitudes for the two configurations in agreement with experimental observation. Since $\sigma^{A}_{xz}$ and $\sigma^{A}_{zy}$ keep the sign of their normal counterparts, the chirality is expected to be identical for the two configurations (Fig. 1f) in contrast to the configuration scheme proposed by early studies of polarized neutron diffraction (Fig. 1e).

Thus, the theoretically-expected  and the experimentally-observed AHC are close to each other. But, the strong temperature dependence raises a fundamental question.  Low-field ordinary Hall conductivity is proportional to the square of the mean-free-path\cite{Ong1991,Behnia2016}. On the other hand, topological Hall conductivity, like in Quantum Hall Effect\cite{Kohmoto1985}, does not depend on the mean-free-path. If the AHE is a topological property of the electronic system and independent of the carrier-mean-free-path, why does its magnitude decrease by a factor of three when the system is warmed from 200K to 400K? There is no trace of such a strong variation in iron\cite{supplement}. The magnitude and the temperature dependence of ANE brings additional insight. Ordinary Nernst response roughly scales with k$_{B}$T/E$_{F}$ \cite{Behnia2016}. Here, with cooling, $\alpha^{A}_{ij}$ rises faster than $\sigma^{A}_{ij}$ (See Fig. 3).  The $\alpha^{A}_{ij}/\sigma^{A}_{ij}$  ratio evolves from 15 $\mu V/K$ at room temperature  to 50 $\mu V/K$ around 220K. This is a sizeable fraction of k$_{B}$/e $= 86 \mu V/K$ and, unlike its ordinary counterpart, is not attenuated by k$_{B}$T/E$_{F}$.  A $\alpha^{A}_{ij}/\sigma^{A}_{ij}$ ratio close to k$_{B}$/e  appears to confirm the purely topological origin of AHE. The temperature dependence  may be due to a temperature-induced shift in the locus of Weyl nodes.

In summary, we quantified the anomalous transverse response of Mn$_{3}$Sn. The Wiedemann-Franz law is robust with no sign of downward deviation, expected and observed in presence of inelastic scattering. This implies that: i)the electrical Hall current has a purely thermal counterpart carried by Fermi surface quasi-particles; ii) in  Mn$_{3}$Sn, in contrast to common ferromagnets, inelastic scattering does not play any role in generating the AHE response, points to a purely topological origin.

We acknowledge useful discussions with M. O. Goerbig. Z. Z. was supported by the 1000 Youth Talents Plan and the work was supported by the National Science Foundation of China (Grant No. 11574097) and The National Key Research and Development Program of China(Grant No.2016YFA0401704). K. B was supported by China High-end foreign expert programme,111 program (B13033) and Fonds-ESPCI-Paris.

\noindent * \verb|zengwei.zhu@hust.edu.cn|\\
* \verb|kamran.behnia@espci.fr|\\

\appendix
\pagebreak
\section{Samples and experimental methods}
The  Mn$_{3}$Sn single crystal was grown using the Bridgman-Stockbarger technique. The raw materials, Mn (99.99\% purity) and Sn (99.999\%), were weighed and mixed inside an Ar glove box with a molar ratio of 3.1:1. After being sealed in a quartz ampule , they were loaded into an alumina crucible. The growth temperature was controlled at the bottom of the ampule. The material was heated up to 1100 $^{\circ}$C. remained there for 2 hour to ensure homogeneity of the melt, and was cooled slowly to 950 $^{\circ}$C. It was annealed at that temperature for 40 hours afterwards. Finally, it was gently cooled down back to room temperature. The single crystals were cut by a wire saw into $0.5\times0.5\times2mm^3$ dimensions for transport measurements. The iron single crystal with 99.99\% purity was obtained commercially.

The thermal conductivity measurements were done with a one-heater-two-thermometers technique in PPMS with a high-vacuum environment. Two Chromel-constantan (type E) thermocouples were employed to measure the temperature difference generated by a small heater chip. We measured the other transport coefficients separately from thermal conductivity measurement. The voltage was monitored by DC-nanometers (Keithley 2182A) and electric current was driven by a current source (Keithley 6221). Generally, the temperature gradient is around 5 K/cm. The procedure of zero-field transport coefficients were obtained as the reported\cite{Nakatsuji2015} by cooling down the temperature to 5K with a field 7T(-7T), decreasing to +0T(-0T), then starting to measure at different temperatures. The coefficients were finally gotten by being symmetrized(antisymmetrized) accordingly from the raw data. The magnetization measurement was also carried out in PPMS with VSM option with 0.1T field-cooling.

\section{Carrier density and mean-free-path}
The normal Hall conductivity, extracted from the slope of $\sigma_{xz}(B)$ allows to estimate the Hall coefficient, R$_{H}$. This leads to a carrier density of n=2$\times$ 10$^{22}$cm$^{-3}$, close to what was reported previously\cite{Nakatsuji2015}. With this carrier density, and $\rho (300 K)= 250 \mu\Omega cm$, one finds a room-temperature mean-free-path  of  0.7 nm.

\section{ANE, AHE and ARLE in iron}
Room-temperature field dependence of Hall resistivity and Nernst coefficient in an iron crystal is presented in Fig.S1. The iron crystal was cut by a wire saw from a big crystal oriented as (110) obtained commercially. Then the crystal was polished by a sand paper into a typical dimension 3mm$\times$0.8$\times$0.05mm. Because iron has BCC structure, so we regard it has very small anisotropy. The results are similar to an early study of AHE\cite{Dheer1967} and a very recent study of ANE\cite{Watzman1967} in iron. We find $\sigma^{A}_{xy}(300K)=1200 S/cm$, compared to 1030 S/cm reported by Dheer\cite{Dheer1967} and 755 S/cm according to several theoretical calculations\cite{Gosalbez2015,Yao2004,Wang2006}. The magnitude of $\alpha^{A}_{xy}(300K)$ was found to be 18 mA/K cm.

\begin{figure}
\includegraphics[width=9cm]{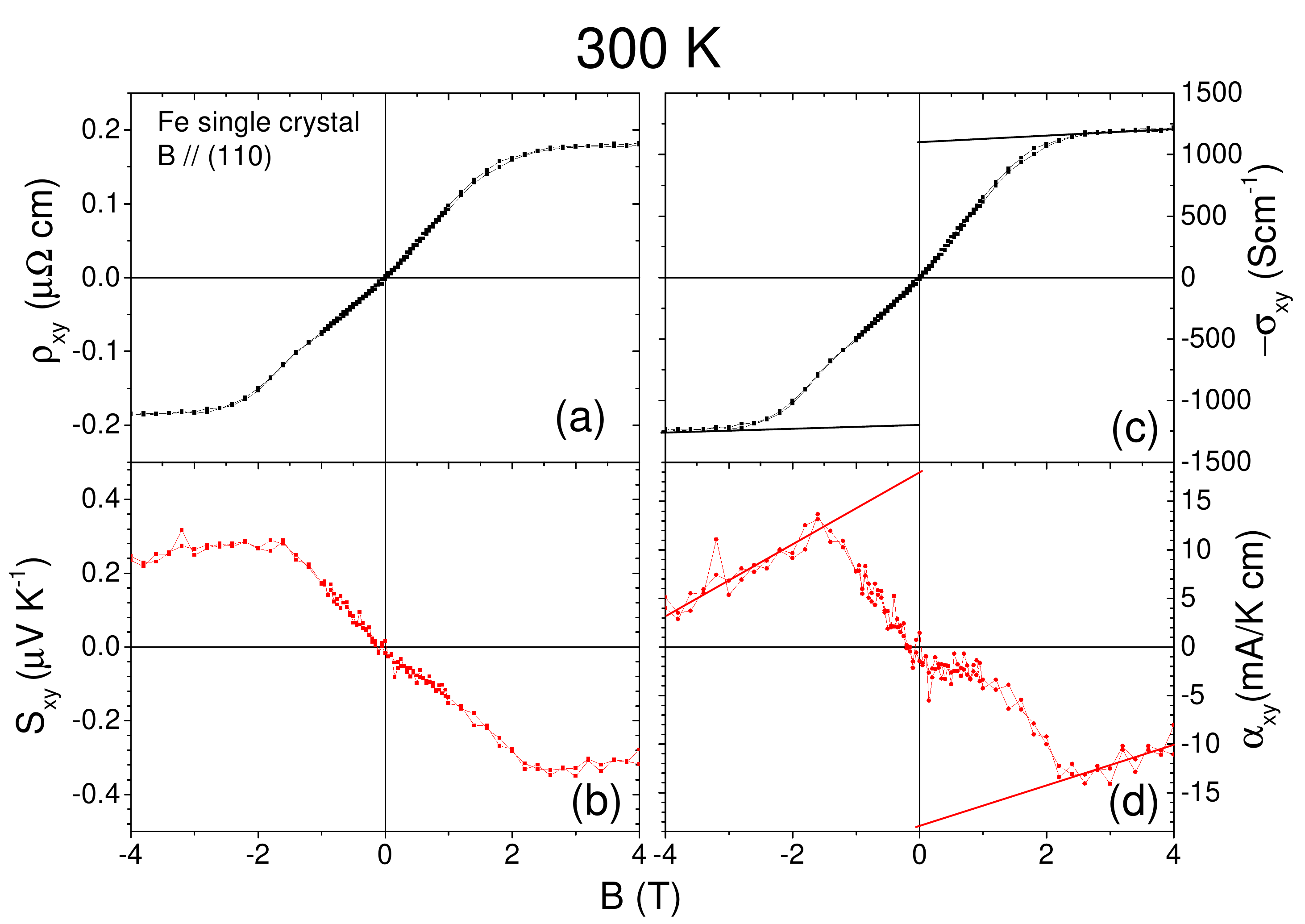}
\caption{  Anomalous Hall effect and Anomalous Nernst effect (bottom) in BCC Iron at room temperature. By extrapolating the slope of the high-field response, one can extract $\sigma^{A}_{xz}(300K)$ and $\alpha^{A}_{xy}(300K)$. }

\end{figure}

We carried out a detailed study of $\sigma_{xy}$ and $\kappa_{xy}$  in another iron single crystal and extracted the normal and anomalous components of the Hall and the Righi-Leduc effects (see Fig. S2). Our results agree with what was reported by Shiomi \emph{et al.}\cite{Shiomi2010}.  One can see that $\sigma^{A}_{xy}$ shows a modest variation with temperature and remains close to the theoretically-expected value.

\begin{figure*}
\centering
\includegraphics[width=15cm]{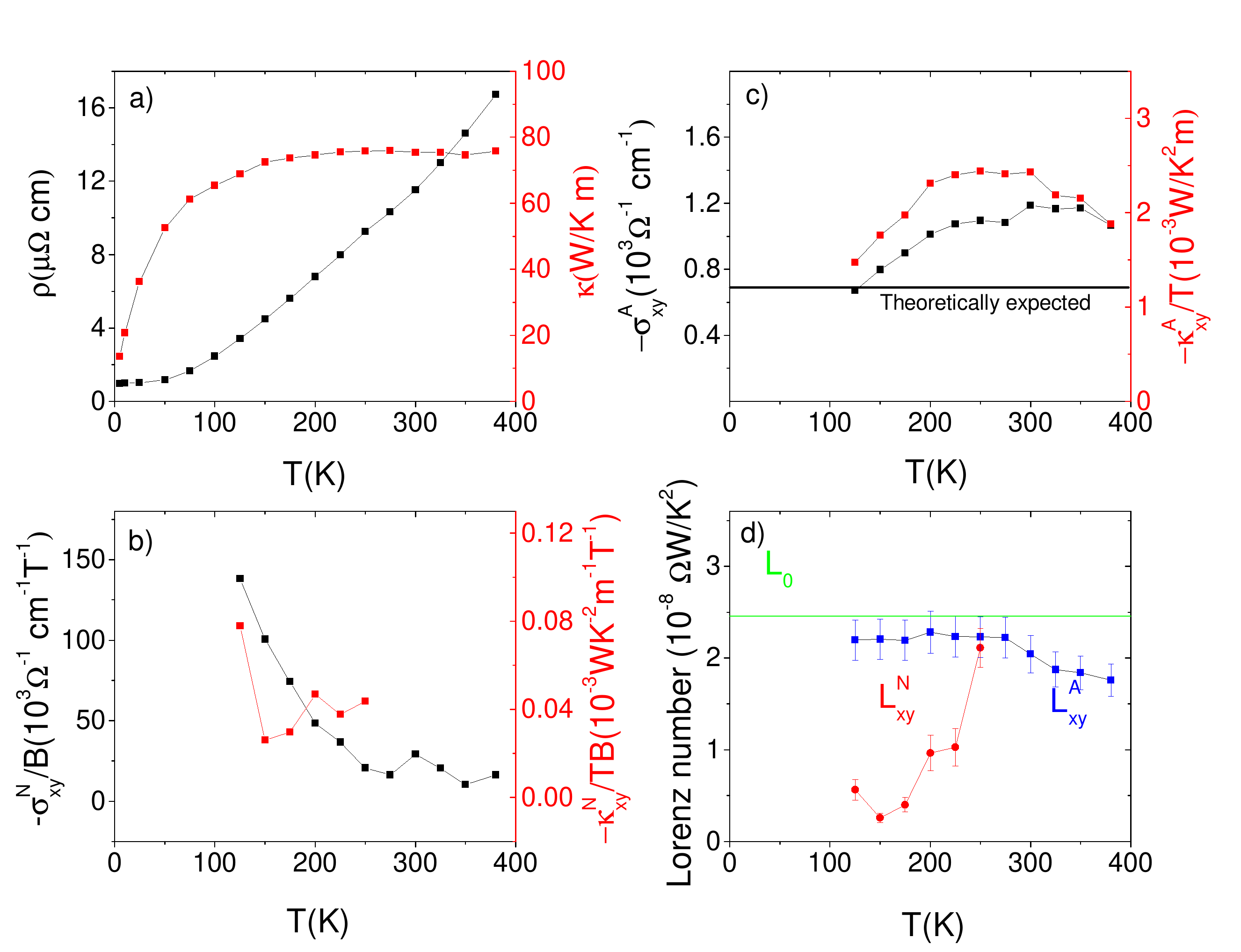}
\caption{ a) Temperature dependence of resistivity and thermal conductivity in an iron crytsal. b) Temperature dependence of ordinary Hall and Righi-Leduc conductivities. c)Temperature dependence of anomalous Hall and Righi-Leduc conductivities. The solid horizontal line represents the theoretically expected value. d) The extracted magnitude of anomalous and ordinary transverse Lorenz-Sommerfeld ratio. As found in a previous study\cite{Shiomi2010}, the anomalous component is much less affected by inealstic scattering}
\end{figure*}

\section{Field dependence of the transverse thermal and electric conductivity in Mn$_{3}$Sn: raw data}
The raw data at different temperatures  is presented here. Fig. S3 shows the field-dependence of $\kappa_{xz}$ and $\sigma_{xz}$ , $\kappa_{zy}$ and $\sigma_{zy}$ at different temperatures, which has been used to plot Fig.4 in the main text. The large error margin at 380K is due to a long waiting time and unavoidable temperature drift.

\begin{figure*}
\centering
\includegraphics[width=18cm]{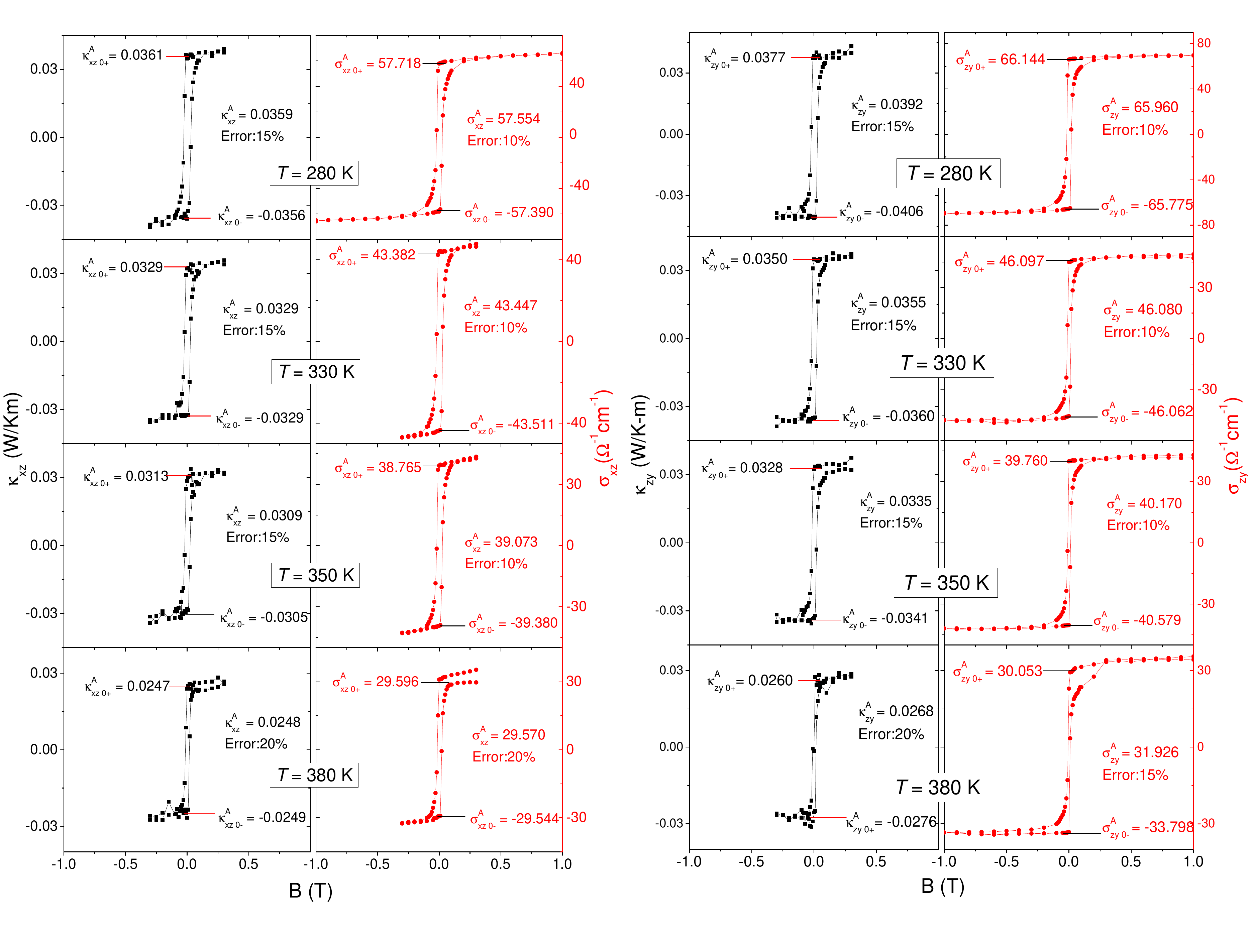}

\caption{The field-dependence of $\kappa_{xz}$ and $\sigma_{xz}$, $\kappa_{zy}$ and $\sigma_{zy}$ at different temperature.}
\end{figure*}

\section{Off-diagonal components of $\overline{\sigma}$, $\overline{\alpha}$ and $\overline{\kappa}$}

The off-diagonal component of the Peltier-Ettingshaussen tensor, $\overline{\alpha}$, is linked to the charge current density, $\overrightarrow{J_{e}}$, the electric field, $\overrightarrow{E}$,  the thermal gradient, $\overrightarrow{\nabla T}$ and the electric conductivity tensor, $\overline{\sigma}$, through:

\begin{equation}\label{1}
 \overrightarrow{ J_{e}}=\overline{\sigma}\overrightarrow{E}- \overrightarrow{\nabla T}\overline{\alpha}
\end{equation}

In absence of temperature gradient, $\nabla T=0$, the tensor $\overline{\sigma}$ can be obtained from measurable resistivity tensor $\overline{\rho}$:

\begin{equation}
\overline{\sigma}=\overline{\rho}^{-1}.
\end{equation}

In absence of charge current density, one finds that the  Seebeck-Nernst tensor $\overline{S}$ is linked to $\overline{\alpha}$ and $\overline{\sigma}$ in the following manner:

\begin{equation}\label{2}
\overline{S} =\overline{\sigma}^{-1} \overline{\alpha}= \overline{\rho} \overline{\alpha}
\end{equation}

Therefore, combining the measured components of $\overline{S}$ with those of the resistivity tensor, $\overline{\rho}$,  one can compute the components of $\overline{\alpha}$.  Specifically, for sample 1: when the magnetic field along y axis([01\={1}0] in the our case), the electric and thermal current along z axis[0001] and the transverse voltage along x axis[2\={1}\={1}0].\\
When the electric field is applied along z axis:
\[
\begin{array}{cc}
\overline{\sigma}=\overline{\rho}^{-1}=
\left( \begin{array}{cc}
\rho_{xx} &\rho_{xz} \\
\rho_{zx} &\rho_{zz}
\end{array} \right)^{-1}
\\\\
=\frac{1}{\rho_{xx}\rho_{zz}-\rho_{xz}\rho_{zx}}
\left( \begin{array}{cc}
\rho_{zz} &-\rho_{xz} \\
-\rho_{zx} &\rho_{xx}
\end{array} \right)=
\left( \begin{array}{cc}
\sigma_{xx} &\sigma_{xz} \\
\sigma_{zx} &\sigma_{zz}
\end{array} \right)
\end{array}
\]
The Hall conductivity can be deduced as the following:
\begin{eqnarray*}
\sigma_{xz}=\frac{-\rho_{xz}}{\rho_{xx}\rho_{zz}-\rho_{xz}\rho_{zx}}.
\end{eqnarray*}
Since $\rho_{xz}=-\rho_{zx}$,and $\rho_{xz}^{2}$ is negligibly small comparing to the longitude resistivity in the current case.
\begin{equation}
\sigma_{xz}\approx\frac{-\rho_{xz}}{\rho_{xx}\rho_{zz}}.
\end{equation}

When there is a temperature gradient, but no electric current along the z-axis:
\[
\begin{array}{cc}
\overline{\alpha}=\overline{\rho}^{-1}\overline{S}=
\left( \begin{array}{cc}
\rho_{xx} &\rho_{xz} \\
\rho_{zx} &\rho_{zz}
\end{array} \right)^{-1}
\left( \begin{array}{cc}
S_{xx} &S_{xz} \\
S_{zx} &S_{zz}
\end{array} \right)
\\\\=\frac{1}{\rho_{xx}\rho_{zz}-\rho_{xz}\rho_{zx}}
\left( \begin{array}{cc}
\rho_{zz} &-\rho_{xz} \\
-\rho_{zx} &\rho_{xx}
\end{array} \right)
\left( \begin{array}{cc}
S_{xx} &S_{xz} \\
S_{zx} &S_{zz}
\end{array} \right)
\\\\=\frac{1}{\rho_{xx}\rho_{zz}-\rho_{xz}\rho_{zx}}
\left( \begin{array}{cc}
\rho_{zz}S_{xx}-\rho_{xz}S_{zx} &\rho_{zz}S_{xz}-\rho_{xz}S_{zz} \\
-\rho_{zx}S_{xx}+\rho_{xx}S_{zx} &-\rho_{zx}S_{xz}+\rho_{xx}S_{zz}
\end{array} \right)
\\\\=
\left( \begin{array}{cc}
\alpha_{xx} &\alpha_{xz} \\
\alpha_{zx} &\alpha_{zz}
\end{array} \right)
\end{array}
\]
So the off-diagonal term can be written as following:
\begin{equation}
\alpha_{xz}=\frac{\rho_{zz}S_{xz}-\rho_{xz}S_{zz}}{\rho_{xx}\rho_{zz}-\rho_{xz}\rho_{zx}}\approx\frac{\rho_{zz}S_{xz}-\rho_{xz}S_{zz}}{\rho_{xx}\rho_{zz}}.
\end{equation}

Using the measured known components of the resistivity and the Seebeck-Nernst tensor,  $\alpha_{xz}$ can be obtained.

To obtain the thermal Hall coefficient, the thermal current can be written as following when only temperature gradient is present:
\begin{eqnarray*}
\overrightarrow{J_{q}}=-\overline{\kappa}\overrightarrow{\nabla T}
\end{eqnarray*}
\[
 \left( \begin{array}{cc}
J_{qx} \\
J_{qz}
\end{array} \right)=-
\left( \begin{array}{cc}
\kappa_{xx} &\kappa_{xz} \\
\kappa_{zx} &\kappa_{zz}
\end{array} \right)
\left( \begin{array}{cc}
\partial_x T \\
\partial_z T
\end{array} \right)
\]
Because thermal current density along z axis, we obtain:
\[
0=-\kappa_{xx}\partial_x T-\kappa_{xz}\partial_z T
\]
\[
J_{qz}=-\kappa_{zx}\partial_x T-\kappa_{zz}\partial_z T
\]
So the thermal current density can be obtained:
\[
J_{qz}=\frac{\bigtriangledown_x T}{\kappa_{xz}}(\kappa_{xx}\kappa_{zz}+\kappa_{xz}^2)
\]
Since $\kappa_{xz}\ll\kappa_{xx}$ or $\kappa_{zz}$ , we can drop the $\kappa_{xz}^2$ term. With $J_{qz}=I^2R/(wt)$, where $t$, $w$ is the thickness, width of the sample respectively. The Righi-Leduc coefficient can be inferred:

\begin{equation}
\kappa_{xz}=\frac{\partial_x T \kappa_{xx}\kappa_{zz}}{J_{qz}}=-\frac{\kappa_{xx}(B)\kappa_{zz}(B)\Delta_x T(B)t}{I^2R}
\end{equation}

The same procedure can be used for a second configuration: the magnetic field is along x axis, the electric and thermal current along y axis and the transverse voltage along z axis . One would derive  the following expressions for $\sigma_{zy}$,  $\alpha_{zy}$ and $\kappa_{zy}$:

\begin{equation}
\sigma_{zy}\approx\frac{-\rho_{zy}}{\rho_{yy}\rho_{zz}}.
\end{equation}

\begin{equation}
\alpha_{zy}\approx\frac{\rho_{yy}S_{zy}-\rho_{zy}S_{yy}}{\rho_{yy}\rho_{zz}}.
\end{equation}
\begin{equation}
\kappa_{zy}=-\frac{\kappa_{yy}(B)\kappa_{zz}(B)\Delta_z T(B)t_2}{I^2R}
\end{equation}

\end{document}